\documentclass[a4paper,11pt]{article}
\pdfoutput=1
\usepackage{pos}

\title{Perturbative Charm Production and the Prompt Atmospheric Neutrino Flux in light of RHIC and LHC}
\ShortTitle{Prompt atmospheric neutrino flux}

\author*[a]{Atri Bhattacharya}
\author[b]{Rikard Enberg}
\author[c]{Mary Hall Reno}
\author[d,e]{Ina Sarcevic}
\author[f]{Anna Stasto}

\affiliation[a]{Space sciences, Technologies and Astrophysics Research (STAR) Institute, Université de Liège,\\
                Bât.~B5a, 4000 Liège, Belgium}

\affiliation[b]{Department of Physics and Astronomy, Uppsala         
                University,\\Box 516, Uppsala, 751 20 Sweden}

\affiliation[c]{Department of Physics and Astronomy,
                University of Iowa,\\30 North Dubuque Street, Iowa City, IA, 52242 U.S.A.}
\affiliation[d]{Department of Physics, University of Arizona,\\1118 E. Fourth Street, Tucson, AZ, 85721 U.S.A.}
\affiliation[e]{Department of Astronomy and Steward Observatory, University of Arizona,\\933 North Cherry Avenue, Tucson, AZ, 85721 U.S.A.}
\affiliation[f]{Department of Physics, Pennsylvania State University,\\University Park, PA 16802, U.S.A}
\emailAdd{A.Bhattacharya@uliege.be}
\emailAdd{rikard.enberg@physics.uu.se}
\emailAdd{mary-hall-reno@uiowa.edu}
\emailAdd{ina@physics.arizona.edu}
\emailAdd{astasto@phys.psu.edu}

\abstract{Prompt neutrinos due to the decay of charmed mesons
          produced in the atmosphere from 
          cosmic-ray and atmospheric nuclei interactions may be
          a significant source of background to ultra-high energy neutrino searches above 10 TeV.
          We re-evaluate this flux using updated charm production cross-sections
          based on QCD parameters, the charm quark mass, and the range for the factorization
          and renormalization scales that provide the best description of this data at fixed target experiments, at RHIC, and at LHC.
          We find that the prompt neutrino flux
          is reduced from previous results in
          the literature by a factor between two and eight, depending on the energy.
          We discuss the implications of our results for current IceCube data.}

\FullConference{%
  40th International Conference on High Energy physics - ICHEP2020\\
  July 28 - August 6, 2020\\
  Prague, Czech Republic (virtual meeting)
}

\usepackage{graphicx}
\usepackage{bm}
\usepackage{caption}
\usepackage{subcaption}
\usepackage{paralist}
\usepackage{multirow}
\usepackage{wrapfig}

\usepackage[all]{hypcap}

\newcommand{\be}{\begin{eqnarray}}
\newcommand{\beq}{\begin{equation}}
\newcommand{\eeq}{\end{equation}}
\newcommand{\ee}{\end{eqnarray}}

\newcommand{\bmp}{\noindent\begin{minipage}{16cm}}
\newcommand{\emp}{\end{minipage}\vskip 7mm} 

\newcommand{\nue}{\ensuremath{\nu_{e}}}
\newcommand{\numu}{\ensuremath{\nu_{\mu}}}

\newcommand{\eg}{\textit{e.g.}}
\newcommand{\ie}{\textit{i.e.}}

\newcommand{\diff}{\ensuremath{\mathrm{d}}}

\begin{document}
\maketitle

\section{Introduction}
Cosmic ray protons incident at the earth with energies exceeding hundreds of
TeV interact with atmospheric nuclei leading to the production of mesons.
These thereafter decay and produce neutrinos that form the
dominant neutrino flux between energies spanning
tens of GeV to hundreds of TeV.
The lower energy end of this spectrum --- up to a few tens of TeV ---
is dominated by neutrinos
from the decays of light mesons --- pions and kaons ---
produced in the atmosphere (\eg\
$ \pi^{\pm} \to \mu^{\pm} \numu \to e^{\pm} \nue \numu \numu$).
The resulting neutrino flux between energies of 10 GeV--1 TeV has been
observed by several experiments (see \cite{Gaisser:2005dt} for a
review) over the last two decades.
At higher energies, as the production of heavier mesons
such as the charmed mesons ($D^{\pm,0}$) becomes increasingly
favoured kinematically,
the proportion of neutrinos from the decay of these mesons in the total
atmospheric neutrino flux gradually grows.
Charmed mesons have shorter lifetimes in comparison to the lighter
pions and kaons;
therefore they decay promptly, without losing energy
between production and decay, leading to a harder 
spectral shape for the resulting neutrino flux, called the prompt neutrino flux.
At energies of about 100 TeV, this flux starts dominating
over the conventional neutrino population.

The understanding of prompt neutrino production
assumes special relevance in light of recent
successes of IceCube towards detecting ultra-high energy neutrinos
extending from tens of TeV to a few PeV \cite{IceCube, Aartsen:2014gkd}, where it might be
the key background beyond 100 TeV energies.
Unfortunately, this understanding is hampered by
the large uncertainties in the QCD modeling of heavy meson production: $pN \to c\bar{c}X$.
Within the realm of perturbative QCD,
uncertainties arise from from multiple facets:
\begin{inparaenum}[\itshape a\upshape)]
	\item the charm mass ($ m_c $),
	\item the factorization ($ M_F $) and renormalization ($ \mu_R $) scales, and
	\item the choice of parton distribution functions (PDFs).
\end{inparaenum}
Hadronization of charm to produce $D$-mesons may be modeled by
different phenomenological models, adding to the uncertainty.
Finally, computing the prompt lepton flux involves
folding the charmed hadron production cross-section with the nucleon
flux in the incoming cosmic-ray flux, thereby incurring large
uncertainties from the limited understanding of the extremely
high-energy cosmic ray composition.

The previous benchmark for UHE prompt neutrino flux \cite{ers} used the
dipole model for their computation.
Our focus in this talk will be on recomputing this flux
to incorporate improvements in the understanding of parameters describing perturbative charm production thanks to recent collider data and more reliable, state-of-the-art cosmic-ray models.
We will discuss the implications of these revised estimates of
the prompt flux for ultra-high energy experiments like IceCube.

\section{Charm production cross section}
Uncertainties in the parameters describing $pp \to c\bar{c}X$
are now better understood thanks to the measurement of the charm
production cross-section at LHC \cite{Abelev:2012vra,ATLAScharmconf,Aaij:2013mga} and
RHIC \cite{Adare:2006hc,Adamczyk:2012af}.
In addition, modern parton distribution functions are less uncertain
at smaller $x$, in part thanks to collider data at higher and higher energies.
A combination of these factors enables the determination of
the charm production cross section more accurately than ever before.

\begin{figure}[tb]
  \centering
  \includegraphics[width=0.75\textwidth]{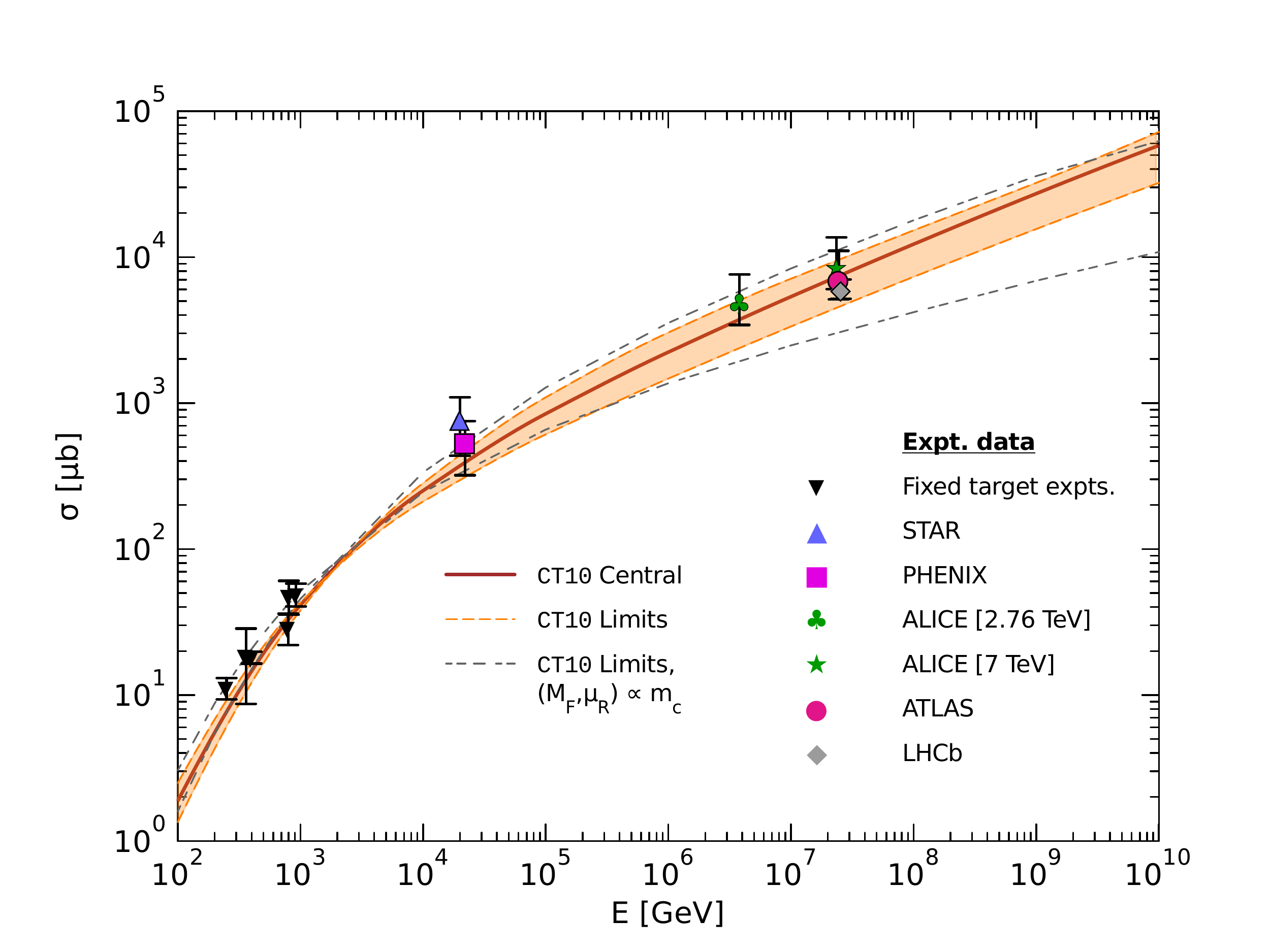}
  \caption{The charm production cross section
           $\sigma_{pN  \to c\bar{c} +X }$ at NLO with $m_c=1.27$ GeV using the
           \texttt{CT10} parton distributions for a range of scales described in the text.
           Also shown are the lower and upper limits (grey fine-dashed curves) when the
           scales are made to vary proportionally to $ m_c $ rather than to $ m_T $.
           For a list of experimental data shown here, see
           Ref.~\cite{Bhattacharya:2015jpa}.}
  \label{fig:sigccb}
\end{figure}

Specifically, we use the range of factorization and renormalization scales determined in Ref.~\cite{Nelson:2012bc} to best describe
current charm production data in colliders: $ M_F/m_c=1.3 $--$ 4.3 $ and $\mu_R/m_c=1.7$--$1.5$ with $M_F=2.1 m_c$ and $\mu_R=1.6m_c$ as the central values when using the \verb+CT10+ PDFs.
We evolve the scales in multiples of the charm transverse mass $m_T^2 \equiv (m_c^2+p_T^2)$, while
using a fixed charm mass of $1.27$ GeV, motivated by lattice QCD results.
The charm production cross section is evaluated to the
next-to-leading-order (NLO) in perturbative QCD
\cite{Cacciari:2001td}.
Fig.~\ref{fig:sigccb} shows the resulting central curve for the charm
production cross section as well as its range of uncertainty
when varying the two scales withing the range described above.

\section{Prompt lepton flux}
The change in the flux of a particle $j$ as it traverses
a slant depth $X$ in the atmosphere is described by the equation:
\begin{equation}
    \frac{\diff \Phi_j}{\diff X}
    = -\frac{\Phi_j}{\lambda_j}   
      - \frac{\Phi_j}{\lambda^\text{dec}_j}
      + \sum_k S(k \to j)\,,
\end{equation}
with flux losses described by the interaction (decay) length $\lambda\ (\lambda^\text{dec})$
and its regeneration from a different particle $k$ described
by the generation function $S (k \to j)$, which is
generically a function of both the particle energy $E_k$ and the slant depth $X$.
To compute the lepton flux from the final neutrino
flux we adopt a semi-analytical approach to solve these
equations using spectrum-weighted $Z$-moments that
depend on the energy alone.
For example, the hadron production moments are defined by
\begin{equation}
Z_{ph}(E_h) = \int_{x_{E_\text{min}}}^1\frac{dx_E}{x_E}
              \frac{\phi_p^0(E_h/x_E)}{\phi_p^0(E_h)}\frac{1}{\sigma_{pA}(E_h)}
              \times
              A\frac{d\sigma}{dx_E}(pN\to hX)\,,
\end{equation}
where, $A = 14.5$ is the average atomic number of target
air nuclei\footnote{We use the approximation $\sigma(pA\to c\bar{c}X)\simeq A\sigma(pN \to c\bar{c}X)$ throughout.},
\begin{equation}
\frac{d\sigma}{dx_E}(pA\to hX)=A\int_{x_E}^{1}
\frac{dz}{z}\frac{d\sigma}{dx_c}(pN\to cX) D_c^h(z)
\end{equation}
in terms of $x_E=E_h/E_b$ and $x_c=E_c/E_b=x_E/z$ for an incident cosmic ray nucleon energy (beam energy) $E_b$,
and fragmentation functions $D_c^h$ assumed to be
energy independent.
The all-nucleon cosmic ray flux is described by
$\phi_p(E,X) \simeq \phi_p^0(E)\exp(-X/\Lambda_p) $.

\begin{figure}[tbh]
  \begin{subfigure}{0.49\textwidth}
    \centering
    \includegraphics[width=0.98\textwidth]{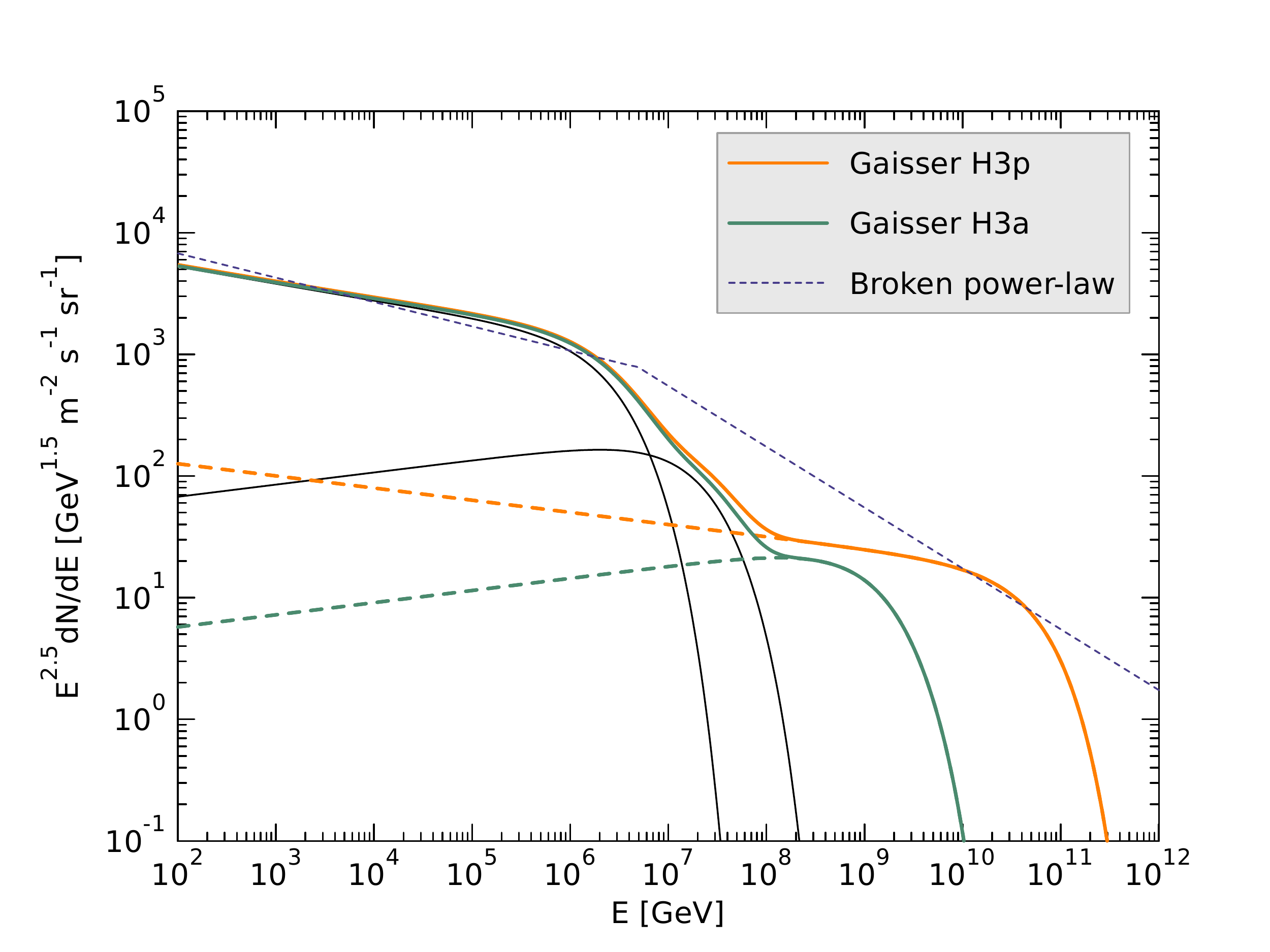}
    \caption{}
    \label{fig:crflx}
  \end{subfigure}
  \begin{subfigure}{0.49\textwidth}
    \centering
    \includegraphics[width=0.98\textwidth]{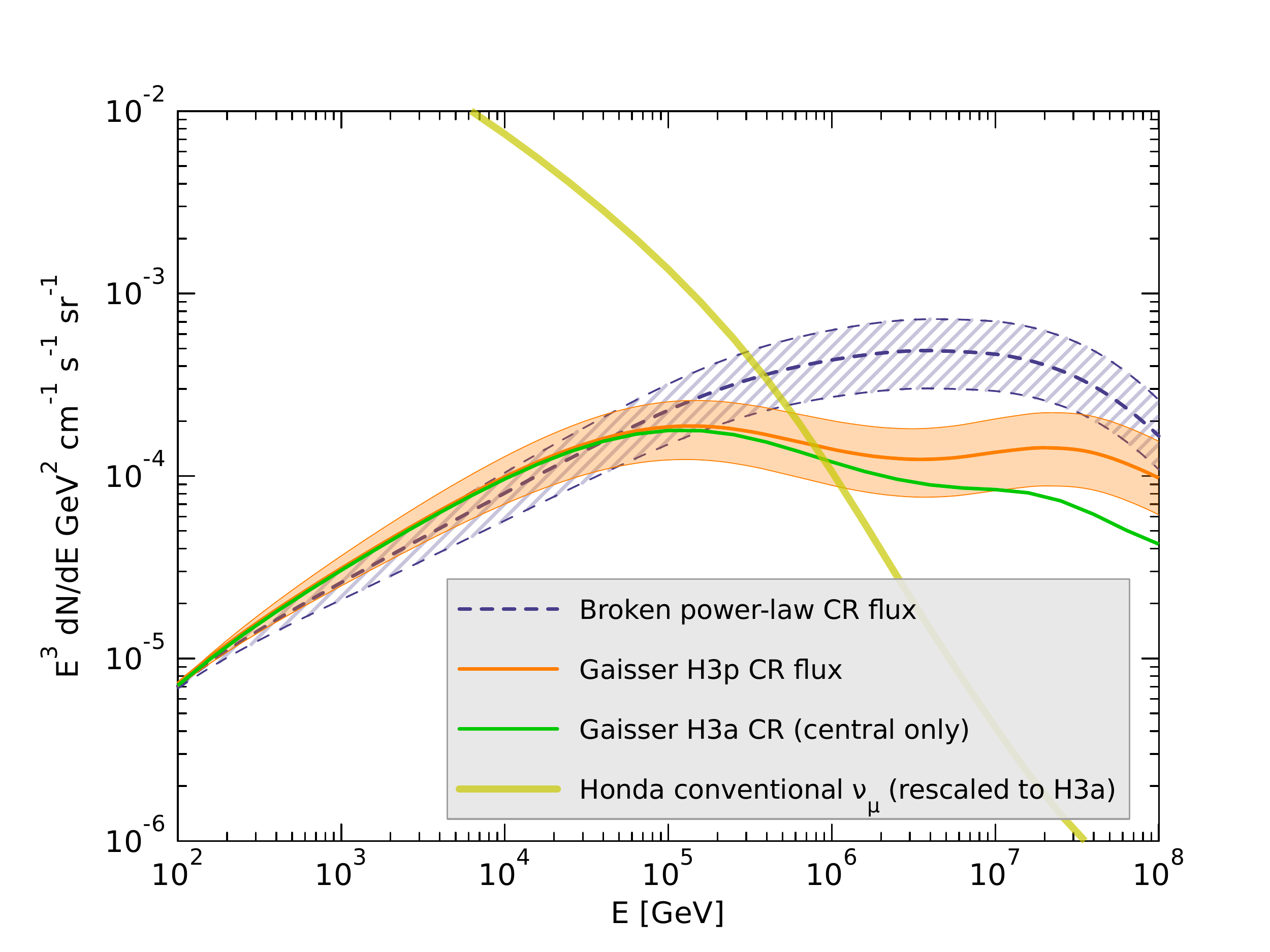}
    \caption{}
    \label{fig:promptiniceflx}
  \end{subfigure}
  \caption{(a) The all-nucleon cosmic ray spectrum as a
               function of energy per nucleon for the three
               component model of ref.~\cite{Gaisser:2012zz}
               with a mixed extragalactic population (H3a)
               and all proton extragalactic
               population (H3p), and for a broken
               power-law.
           (b) Our benchmark results for the prompt                   $ \nu_\mu + \bar{\nu}_\mu $
               flux scaled by $ E^3 $ (orange curve),
               using the H3p cosmic ray flux.
               The blue curve uses instead
               a broken power-law (used in previous analyses,
               \eg, \cite{ers}).
          }
  \label{fig:promptinice}
\end{figure}

Previous results approximated the cosmic ray flux by
a broken-power law \cite{ers};
here, we instead use a more recent parametrization from ref.~\cite{Gaisser:2012zz} with fluxes from three populations: supernova remnants, other galactic sources, and extragalactic sources.
The H3a flux has a mixed composition in the extragalactic population, while we designate the  \emph{H3p flux} as the one where it is all protons (Fig.~\ref{fig:crflx}).

We skip a full description of the procedure for going from the incident cosmic
ray flux to the final prompt neutrino flux using $Z$-moments, and refer the reader to Ref.~\cite{Bhattacharya:2015jpa} instead.
The resulting prompt neutrino fluxes for different choices of incident cosmic-ray flux and QCD parameters are shown in
Fig.~\ref{fig:promptiniceflx}.

\subsection{Comparison to previous results}
In comparison to the fluxes obtained in \cite{ers},
our benchmark results, \ie\ obtained using the Gaisser
H3p cosmic ray flux and central values of the QCD parameters,
are reduced by a factor ranging from 2 at
lower energies (below 100 TeV) to a maximum of about 8 at high energies (at a few PeV).
Roughly quantifying the changes from the main ingredients
in our calculation, we estimate that
\begin{inparaenum}[\itshape a\upshape)]
    \item the use of updated cosmic-ray fluxes leads to a reduction in
          the prompt neutrino flux by about a factor 1.2--3 at energies
          between 100 TeV--1 PeV;
    \item differences in the large $x$ behavior of the dipole and   
          perturbative QCD evaluations of $pA\to c\bar{c}X$, account
          for the relative fluxes differing by a factor of $\sim 1.5$;
          and
    \item re-evaluation of $Z_{pp}$ moments and using updated $p$-Air
          cross section decreases the flux by about 30\%.
\end{inparaenum}

\section{Implications for IceCube}
\begin{figure}[htb]
  \centering
  \includegraphics[width=0.75\textwidth]{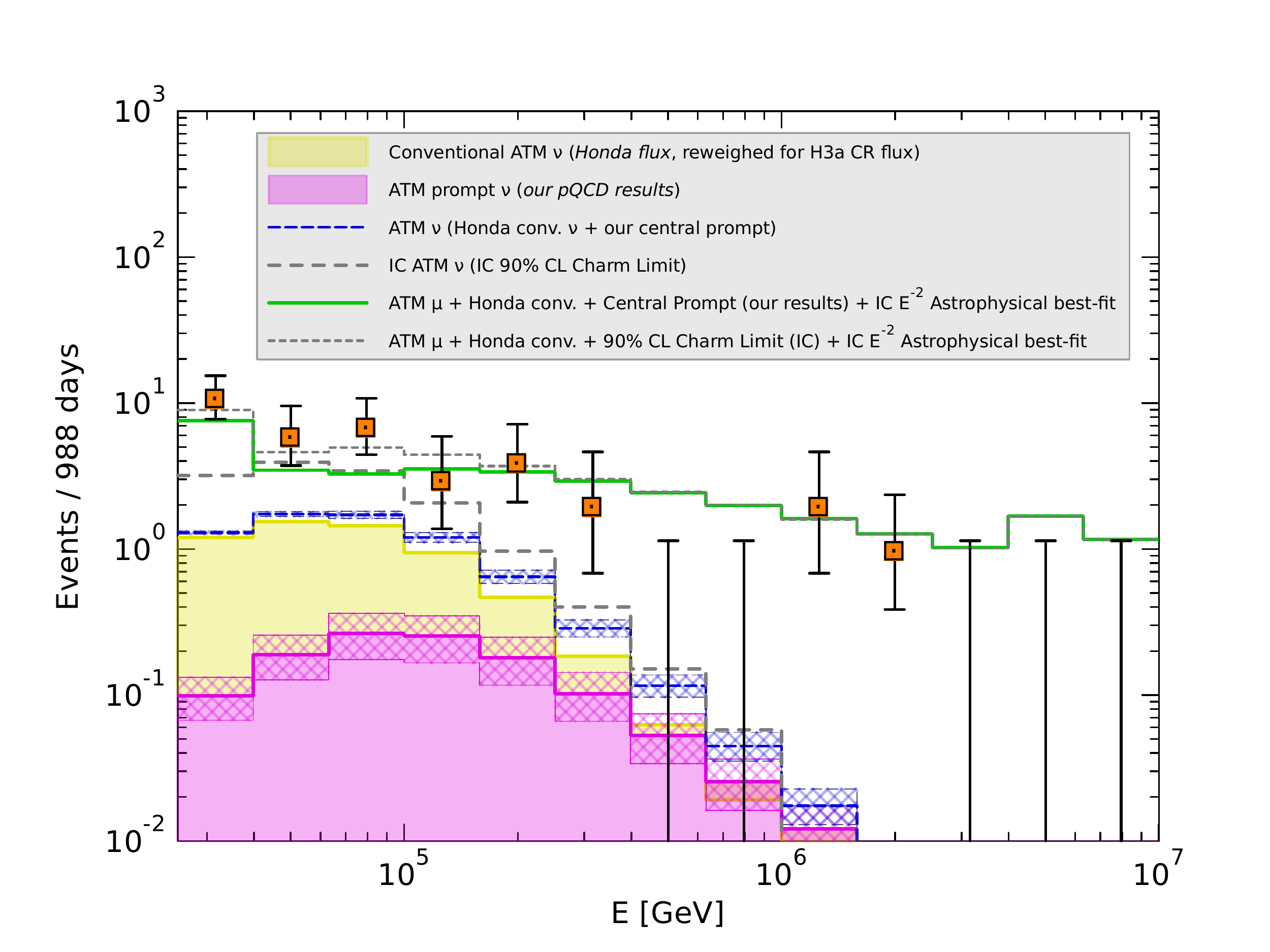}
  \caption{Event rates at IceCube from prompt neutrinos, with our
           updated prediction
           for the prompt flux indicated in magenta, along with uncertainties from variation
           in the QCD parameters indicated as a hatched region around the central curves. See Ref.~\cite{Bhattacharya:2015jpa}
           for details.}
	\label{fig:ICevents}
\end{figure}

Neutrinos produced in the atmosphere, whether conventionally
via pion and kaon decay or via the prompt decay of heavier charmed
mesons, are backgrounds to IceCube's searches for extragalactic
neutrinos at ultra-high energies.
Our calculation revises down existing prompt neutrino flux estimates significantly; consequently, this component of the
background remains subdominant to the conventional flux up to energies of 100 TeV.
Above these energies, the astrophysical flux overwhelms
the atmospheric fluxes anyway;
the prospect for prompt neutrino discovery at IceCube therefore
remains low.
We show the expected break-up of events based on their origins for the IceCube 988-day data \cite{Aartsen:2014gkd}
in Fig.~\ref{fig:ICevents}.

\bibliographystyle{JHEP}
\bibliography{refs}

\end{document}